\documentstyle[aps,prl,multicol,epsfig,amssymb]{revtex}

\def\r{{\mbox{\boldmath$r$}}}

\def\vk{{\mbox{$\hat v$}}}

\def\la{{\langle}}
\def\ra{{\rangle}}

\newcommand{\be}{\begin{equation}}
\newcommand{\ee}{\end{equation}}
\newcommand{\bea}{\begin{eqnarray}}
\newcommand{\eea}{\end{eqnarray}}

\begin{document}
\title{Inverse statistics of smooth signals: the case of two dimensional
turbulence}
 \author{L. Biferale$^1$, M. Cencini$^2$, A. Lanotte$^1$, 
D. Vergni$^2$ and A. Vulpiani$^2$}
\address{$^1$Dipartimento di Fisica and INFM , Universit\`a "Tor Vergata",
Via della Ricerca Scientifica 1, I-00133 Roma, Italy}
\address{$^2$Dipartimento di Fisica and INFM , Universit\`a "La Sapienza",
Piazza Aldo Moro 2, I-00185 Roma, Italy}
\maketitle
\begin{abstract}
The problem of inverse statistics (statistics of distances 
for which the signal fluctuations are larger than a certain threshold)
 in differentiable signals with power law spectrum, 
$E(k) \sim  k^{-\alpha}$, $3 \le \alpha < 5$, is discussed. 
We show that for these signals, with random phases, 
exit-distance moments follow a bi-fractal distribution.
We also investigate two 
dimensional turbulent flows in the direct cascade regime, which display a more
complex behavior.  We give numerical
evidences that the inverse statistics of 2d turbulent flows is described by 
a  multi-fractal probability distribution, i.e. the statistics of laminar
events is not simply captured by the exponent $\alpha$ 
characterizing the spectrum. 
\\
PACS number(s)\,: 47.27.Eq, 05.45.-a
\end{abstract}
%
\begin{multicols}{2}
\noindent 
Many phenomena in natural science posses complex behaviors over
a wide range of spatial and temporal scales. 
Complexity is quantified by the non-Gaussian properties of 
probability distribution functions (pdf) of signal increments 
over a given  range of scales. Whenever  pdf's at different
scales cannot be superposed by a simple rescaling procedure
one speaks about intermittency \cite{Frisch}. Such  non-trivial 
rescaling properties may be exhibited by pdf's tails or peaks, or both
\cite{CLMV01}. 
Strongly intermittent and rough signals, like those encountered 
in three dimensional turbulent flows,  are the typical examples 
of systems with non-trivial, say multiaffine, scaling of pdf's tails. 
Other important natural phenomena develop  simple pdf's tails
 but non-trivial  pdf's peaks. This is the case of  two dimensional
turbulence as it will be shown in this letter.
Indeed laminar fluctuations,
corresponding to the events described by the peak of the 
probability distribution, posses non-trivial scaling properties. 
Recently, it has been shown that laminar fluctuations of rough
and multiaffine fields are optimally characterized in terms
of their exit-distance statistics, also known as inverse-statistics
\cite{mogens,exit-3d,entropyphysicad,entropyphysicad2}.

The aim of this letter is twofold. 
First we want to extend the application
of inverse statistics \cite{mogens,exit-3d} to the case of smooth signals 
with a given power spectrum, $E(k) \sim k^{-\alpha}$. 
In particular, we discuss signals only one time differentiable, 
i.e. $3 \le \alpha <5$.
For such signals,  direct statistics, i.e. moments of signal increments 
over a given scale, do not bring any information: they are always
 dominated by the differentiable event, $v(x+r)-v(x) \sim r$, 
which are always present when the spectrum has a slope
$\alpha>3$. On the contrary, we will show that the exit-distance  
statistics is given by a bi-fractal distribution if only one kind of 
{\it more than smooth}  fluctuation exists. With {\it more than smooth}
fluctuations we mean events where the signal has a local 
scaling as $v(x+r)-v(x) \sim r^h$ with $h>1$.\\
Second, we will apply the inverse statistics analysis
to the case of two dimensional turbulence. Two-dimensional
turbulence in the direct enstrophy
cascade regime is of obvious importance both theoretically and 
practically to understand a variety of different natural processes, e.g.
in geophysics and astrophysics~\cite{lesieur,kraichnan_2d}.
By applying the exit-distance analysis to a set of  
2d numerical simulations, we will show in a quantitative way that 
laminar events of two-dimensional turbulence posses 
highly non-trivial rescaling properties, revealing a rich  
(multifractal) structure of laminar fluctuations.
We also discuss the importance of large-scale structures
in determining the inverse statistics of two dimensional turbulent flows.

Let us begin by considering a one dimensional signal built by 
fixing its spectrum as $E(k) \sim k^{-\alpha}$ on all avai\-la\-ble wave-numbers\,: 
\begin{equation}
v(x) = \sum_k \vk(k) e^{i(xk +\theta_k)}\,,
\label{smooth}
\end{equation}
with $|\vk(k)|^2 \sim k^{-\alpha}$ and $\theta_k$  random phases, 
uniformly distributed in $[0,2\pi]$.
If $3 \le \alpha <5$ the signal is everywhere one-time 
differentiable\,: it follows that moments of its 
difference over any  increment $r$ 
always have a differentiable scaling, namely
$$ S_p(r) = \la (v(x+r)-v(x))^p \ra \sim c_p r^p\,,$$ while 
  moments with  $p\le -1$ do not exist. 
In order to highlight the role of non-trivial
{\it more than smooth}
 stochastic fluctuations associated to the spectrum slope 
$\alpha$, one needs observable sensitive to laminar fluctuations like
 moments of  inverse statistics. 
With inverse statistics we mean moments of increments, 
$r(\delta v)$, necessary to observe
in the signal a forward (backward) exit through a barrier $\delta v$.
In particular, we fix the height of the barrier,
 $ \delta v$, and we
pick at random a reference point $x_0$. Then, we measure 
the first forward or backward  exit, $r(\delta v)$,  i.e. the first point 
  such that  $|v(x_0 \pm r)-v(x_{0})| \ge \delta v$
and we repeat the observations for many $x_0$ and 
for different barrier heights.
This allows us to define the probability distribution
for exit events, ${\cal P}(r(\delta v))$.\\ Let us define, 
the  exit moments through a 
barrier $\delta v$ as:
\begin{equation}
T^{(p)}(\delta v) = \la r^p(\delta v) \ra\,,
\label{exit_mom}
\end{equation} 
where the average is taken with 
respect to the random choice of $x_0$~\cite{note2}. Positive moments
of exit events 
preferentially weight smooth,
 {\it laminar}, fluctuations. Exit-event moments are also called
inverse statistics moments because of the  difference with 
{\it direct} statistics where one measure signal increments over a given
scale.
For the prototype smooth signal (\ref{smooth}), 
a rigorous estimate of the scaling exponents of inverse statistics moments 
can be derived as follows.
From (\ref{smooth}), when the spectrum exponent is in the 
window of first order differentiability, $3 \le \alpha < 5$, we may estimate
the typical fluctuations as
\begin{equation}
v(x_0+r)-v(x_0) \sim \partial_{x_0}v(x_0) \r + c(x_0)r^h \,,
\label{local}
\end{equation}
where we have kept only the two most important scaling behaviors:
$O(r)$ because of the differentiability and $O(r^h)$ from the
spectrum exponent. In (\ref{local}), 
the exponent $1 \le h < 2$ is connected to the
spectrum slope by the  dimensional relation $\alpha=2h+1$, while 
the function $c(x)$ is a continuous function of $x$. By
studying the exit event, in the limit of small barrier height,
we select with probability 
one the differentiable scaling $r(\delta v ) \sim \delta v$ except for 
those  $x_0$s 
where the  first derivative, $\partial_x v(x_0)$, vanishes. 
In the latter  case, the $O(r^h)$ term 
 dominates  the scaling behavior.
With $ 3 \le \alpha < 5$ the first derivative is a self-affine signal
with H\"older exponent $ \xi  = h-1$, i.e. $\partial_x v(x+r) -\partial_x v(x) 
\sim r^{\xi}$, that is it vanishes on a fractal set 
of dimension $D = 1-\xi = 2-h$. Therefore, the probability to see the 
sub-dominant term $O(r^h)$ dominating the exit events   
in (\ref{local}) is given by the probability 
to pick  a point at random on a fractal set with dimension $D$, i.e.
\begin{equation}
P( r \sim (\delta v)^{1/h} ) \sim r^{1-D} = (\delta v)^{1-1/h}.
\end{equation}
Taking into account both situations, we end with the following 
{\it bi-fractal} prediction for inverse statistics  moments\,:
\begin{equation}
T^{(p)}(\delta v) \sim \delta v^{\chi_{\rm bf}(p)},\; \chi_{\rm bf}(p) = 
\min\!\left(p,\frac{p}{h}+1-\frac{1}{h}\right).
\label{bi-frattal}
\end{equation}
From the previous bi-fractal formula,
one sees that laminar, differentiable, fluctuations influence the
inverse statistics only up to moments of order $p=1$; for larger $p$, 
 the pdf is dominated
by the sub-dominant behavior, $v(x+r)-v(x)  \sim r^h$. 
In other words, the extrema of the signal play the role of singularities in 
inverse  statistics\,: close to the extrema, events with much longer
exits through barriers of order $\delta v$
 are observed when $\delta v \rightarrow 0$.
In Fig.~1, we numerically check this prediction on a one-dimensional signal
with $\alpha=4$, i.e. with $h = 1.5$.
The derivative of such a signal has an H\"older exponent $\xi=0.5$, i.e.
it is a stationary Brownian motion. 
As shown, 
the prediction is verified with high accuracy.

Switching to real signals, we consider 
 two dimensional incompressible turbulent flows. 
As it is well known, 2d turbulence is characterized
by a forward enstrophy cascade from forced scales to dissipative 
ones~\cite{kraichnan_2d}.
In the inertial range, arguments {\it \'a la} Kolmogorov give  
for the velocity spectrum the prediction $E(k)\sim k^{-3}$ 
plus logarithmic corrections~\cite{kraichnan_2d}, 
which is experimentally~\cite{exp_2d} and numerically~\cite{borue} 
observed. However, it is also measured a {\it more than
smooth} spectrum $E(k) \sim k^{-\alpha}$ with $\alpha>3$ depending
on the characteristics of the forcing and 
large-scale dissipation~\cite{benzi_2d,ott}.
The dependency of inertial range statistics from large-scale 
effects (universality issue) is still an open problem
in 2d turbulence in the direct cascade regime.
\begin{figure}
\epsfxsize=7.9truecm
\epsfysize=5.5 cm
\epsfbox{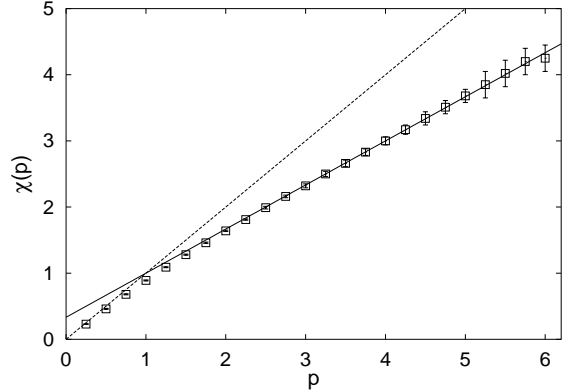}
\caption{Scaling exponents $\chi(p)$ for the $1$-d signal (\ref{smooth}) 
with $\alpha=4$. The dashed line refers to the 
linear differentiable 
behavior for $p \le 1$, $\chi_{\rm bf}(p)=p$. The continuous line
gives  the more than smooth behavior,
 $\chi_{\rm bf}(p)=(p/h+1-1/h)$. Moments (\ref{exit_mom}) have been 
computed using $10^3$ realizations of the signal (\ref{smooth}) 
with $2^{17}$ modes; for each realization  
$2^{12}$ starting points, $x_0$, have been taken at random.}
\end{figure}
\noindent The most important feature of the vorticity cascade is the presence 
of a strong interaction between 
eddies of very different scales\,: such a non-locality (in Fourier space) 
should play a fundamental role in shaping the energy spectrum, in particular 
the strong dependency of the spectrum slope from large-scale effects.
In order to understand the importance of large-scale statistics 
one needs to go beyond the spectrum slope. 

We analyze exit-distance statistics in a series
of direct numerical simulations of the two-dimensional Navier-Stokes equation:
\begin{equation}
\partial_t \omega + J(\omega,\psi)=\nu \Delta^q \omega -\beta_{\rho} \Delta^{-\rho}\omega + F,
\end{equation}
where $\omega$ is the vorticity, $\psi$ the stream function, 
 and $J$  the Jacobian.
We use a standard dealiased pseudo-spectral algorithm with
periodic boundary conditions, at resolutions $512^2$ and $1024^2$.
The large-scale forcing F is Gaussian, white-in-time, 
and nonzero only at some characteristic wave-numbers $k_f$ between 
$4$ and $6$. Enstrophy is dissipated at small scales with 
 an hyper-viscosity $q=4$. Energy is removed
at large scales to avoid piling up on the smallest mode.  We performed 
two sets of numerical simulations with different IR draining, with
an inverse Laplacian with $\rho=2$ case (A) and with 
$\rho=0$ case (B), 
without observing big differences in the spectrum slope.\\  
In Fig.~2 (inset) we show the compensated average spectrum
that we observe for case (A)\,: notice the strong influence of coherent 
structures for low wave-numbers. The best fit to the spectrum
slope for this case gives $\alpha=3.26 \pm 0.06$, i.e. a more than 
smooth exponent $h = 1.13$, while for the case (B) we found 
$\alpha=3.24 \pm 0.06$. 
 \\
The relevant test we wish to perform on the 2d flow
is comparing the inverse  statistics measured on several snapshots
of the direct numerical simulations with the inverse  statistics
obtained after randomization of all velocity 
phases on the same frames. We measured  the  moments of exit events  
using  both transversal and 
longitudinal velocity increments because they could  reflect 
in a different way the presence of coherent structures.
\begin{figure}
\epsfxsize=8.4truecm
\epsfysize=5.5 cm
\hspace{-0.5truecm}
\epsfbox{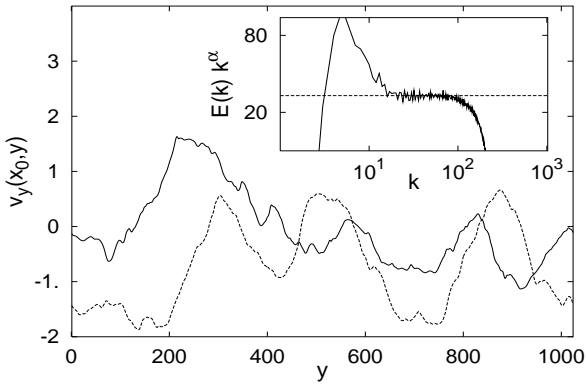}
\caption{One dimensional longitudinal cut of the velocity field $v_y(x_0,y)$
(continuous line); the same after randomization of the phases
(dashed line). In the inset the averaged compensated spectrum, with 
the best fit in the inertial range $\alpha=3.26\pm 0.06$}
\end{figure}
 The rationale for this test is to investigate the importance
of correlations between fluctuations at different wave-numbers
and therefore the ``information'' content brought by 
coherent structures in 2d turbulent flows. 
In Fig.~2, a one-dimensional cut of the 2-d velocity field
before and after phases randomization is plotted. 
At a first glance, it is rather difficult to 
distinguish between the true dynamical and the randomized field. This is
due to the steepness of the  spectrum, i.e. only few modes 
dominate the real-space configuration.
Despite the apparent similarity, there are big
statistical differences between the two fields. Looking at inverse moments
we measure a clear departure of the true turbulent 
 statistics from the bi-fractal prediction (\ref{bi-frattal}), 
while the randomized
configurations are in agreement with it,  with $h=1.13$. 
Due to the numerical limited resolution, in order to perform a 
quantitative statement, one can evaluate only 
relative scaling properties. Therefore, we measure scaling laws of 
the inverse  statistics by plotting all moments  
$T^{(p)}(\delta v)$ versus a reference one, say $T^{(2)}(\delta v)$.
This is the same technique called ESS~\cite{ess} fruitfully applied
in the direct analysis of 3d turbulent data with the aim of re-adsorbing
some finite size effects and extracting scaling
information also at moderate resolution. 
Plotting data in this way allows quantitative statements only for
relative exit-distance exponents, $\chi(p)/\chi(2)$.
\begin{figure}
\epsfxsize=7.9truecm
\epsfysize=5.5 cm
\epsfbox{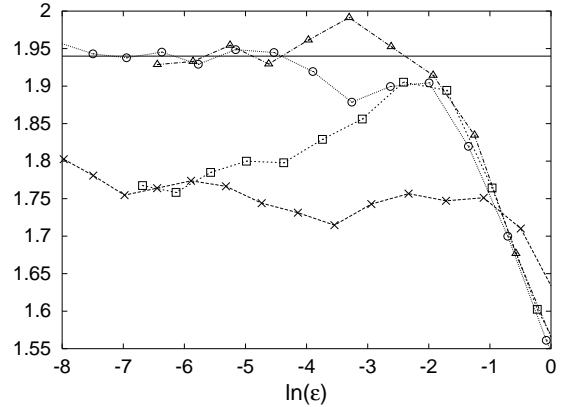}
\vspace{0.2truecm}
\caption{Logarithmic local slopes  for 
the relative scaling $T^{(4)}(\delta v)$ vs 
$T^{(2)}(\delta v)$\,: for the longitudinal ($\Box$) and transverse 
exit-distance ($\times$) for run (A). The same quantities after
the phases randomization are represented by   ($\triangle$) and
 ($\circ$), respectively.}
\end{figure}
In Fig.~3 we show the logarithmic local slopes for the  
relative scaling of $T^{(4)}(\delta v)$ vs $T^{(2)}(\delta v)$ for
longitudinal and transversal velocity increments.
 We stress two important results. 
First, the inverse-statistics moment exponents measured
on the fields from the numerical simulations after phases randomization
coincide with the prediction (\ref{bi-frattal})
with  $h=1.13$ as extracted 
from the averaged spectrum. On the other hand,
the longitudinal and transversal  inverse-statistics moments 
 without phases randomization have a 
more complex, {\it intermittent}, distribution 
 i.e. they are not described by the bi-fractal prediction 
(\ref{bi-frattal}). Second, longitudinal
and transversal  moments are slightly different, indicating 
that longitudinal and transversal velocity fluctuations probe
differently the smooth part of the 2d field. This is of course
connected to the fact that longitudinal or transversal 
velocity differences  have different profiles 
when measured across coherent vortical structures.
We also note that transversal exit moments display a better
scaling behavior than the longitudinal ones.\\
In Fig.~4, we summarize our results showing the curve 
$\chi(p)/\chi(2)$ for both randomized and not-randomized 
longitudinal and transversal exit moments for run (A)~\cite{foot}.  
Notice that for $ p<1$ randomized
and not randomized data almost coincide because those 
moments are dominated by the differentiable fluctuations 
$v(x+r)-v(x) \sim r$ and therefore the relative
scaling exponents differ only due to the factor $\chi(2)$ which
is almost the same for both data sets.  On the other hand, clear different
statistics are measured, for $p>1$, by comparing the  true dynamical 
longitudinal and transversal exit moments with the randomized
ones.

In conclusion, we have given an estimate of inverse-statistics 
moments for signals with a more than smooth spectrum, i.e.
signals which are differentiable and  with non-trivial stochastic
sub-leading fluctuations. We have also shown that statistical
properties of a 2d turbulent flow are not simply summarized by  the 
spectrum slope. 
\begin{figure}
\epsfxsize=7.9truecm
\epsfysize=5.5 cm
\epsfbox{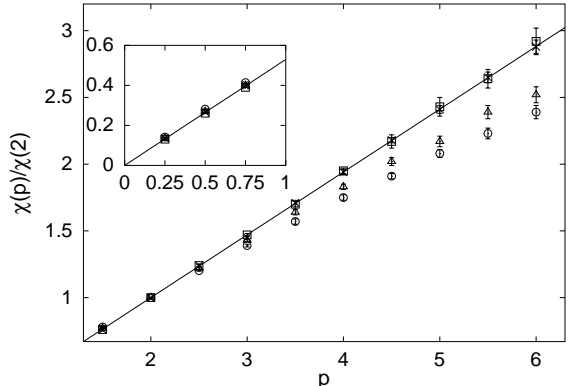}
\caption{$\chi(p)/\chi(2)$ for
 $1.5 \le p \le6$ for both longitudinal ($\triangle$) and 
transversal ($\circ$) 
exit moments obtained in run (A), ($\Box$) and ($\times$) indicate the same
 after randomization, respectively. Inset: the same but for $ p <1$. 
 The continuous line is the bi-fractal prediction 
$\chi_{\rm bf}(p)/\chi_{\rm bf}(2)$,
with $h=1.13$.
 Error bars have been estimated as the maximum 
deviation of the local slopes from the average value.
  Run (A) gives indistinguishable results (not shown).}
\end{figure}
From the  exit-distance analysis it is
 possible to highlight a whole 
spectrum of {\it more than smooth} fluctuations. Such fluctuations, being
connected with laminar events, are the strongest statistical signature
of large scale coherence. Experiments with different methods of 
removing/pumping energy at large scales should be performed, to investigate 
the importance of large-scales structures in the inverse statistics
of flows with different spectra. A more quantitative comprehension of 
the multifractality of inverse statistics could be inferred trying to connect
scaling exponents to the finite time (Lagrangian) Lyapunov exponents
and the drag coefficient, by extending the analysis proposed by 
Ott and collaborators \cite{ott}. 
 As a final remark we would like to stress that
     inverse statistics provide a completely new statistical indicator
     with respect to the standard direct statistics observable.  We
     have shown that such method is necessary in all those cases where
     non-trivial fluctuations are sub-leading with respect to the
     differentiable contributions. Obviously, the same kind of
     analysis here reported can be extended to temporal signals,
     opening the possibility of applying the method to a broad class of
     natural phenomena. As an example, we just mention possible
     applications in situations common to climatology or meteorology
     where estimating the probability of persistent velocity
     configuration, or of any other dynamical variable, is relevant.
     As a perspective, one important generalization would the investigation 
     of multi-dimensional signals by studying the statistics 
     of d-dimensional volumes between equispaced iso-surfaces. 
     The latter method may be, for example, important for analyzing 
     coherence properties of two-dimensional or multi-dimensional patterns.

We acknowledge useful discussions
with R.~Benzi, G.~Boffetta and G.~Eyink. 
This work has been partially supported by the EU under the Grant
No. HPRN-CT  2000-00162 ``Non Ideal Turbulence''   and  the Grant 
ERB FMR XCT 98-0175 ``Intermittency in Turbulent Systems''.
We also acknowledge INFM  support (Iniziativa di Calcolo Parallelo).

\end{multicols}

\begin{thebibliography}{99}
\bibitem{note2} One could also perform a sequential
average, scanning consecutively the time series. 
This is the original procedure used in the calculation of exit distances
for three dimensional turbulence in~\cite{exit-3d}. It is easy to show,
that the sequential procedure is equivalent 
to the random one after a  
 suitable normalization~\cite{exit-3d}.
\bibitem{foot} We  refrain from giving a measurement
for the relative exponent with $p=1$. This is because already in the exactly
solvable case of exit distances of a twice differentiable extremum,
i.e. the case of a signal like (\ref{bi-frattal}) with $h=2$, one can 
prove that for $p=1$ exit-distance moment has logarithmic corrections
which strongly perturb the scaling.
\end{thebibliography}
\end{document}